\documentclass[aps,prx,reprint,floatfix,superscriptaddress,footinbib,twocolumn]{revtex4}
\usepackage{xcolor}
\usepackage{microtype}

\usepackage[bookmarks=true,
bookmarksnumbered=false, 
bookmarksopen=false, 
colorlinks=true,
linkcolor=blue]{hyperref}
\definecolor{webblue}{rgb}{0, 0, 0.5} 
\hypersetup{colorlinks=true,urlcolor=blue,citecolor=blue}
\usepackage{graphicx}
\usepackage{subfigure}
\usepackage{url}
\usepackage{hyperref}
\usepackage{inconsolata}
\usepackage{amsmath}
\allowdisplaybreaks
\usepackage[capitalize]{cleveref}
\usepackage{braket}
\usepackage{outlines}
\usepackage{enumitem}
\usepackage{soul}
\usepackage{color}

\usepackage{bm} 
\DeclareMathOperator{\Tr}{Tr}
\usepackage[utf8]{inputenc}
\usepackage{siunitx,booktabs}
\usepackage{multirow}
\usepackage{amssymb}

\usepackage{scalerel}
\usepackage{relsize}

\usepackage{soul}
\usepackage{tabularx}

\usepackage{amsmath}
\usepackage{amssymb}
\usepackage{bbm}
\usepackage{braket}
\usepackage{xcolor}

\usepackage{comment}
\usepackage{pifont}

\begin{document}

\author{Harley D. Scammell}
\affiliation{School of Mathematical and Physical Sciences, University of Technology Sydney, Ultimo, NSW 2007, Australia}

\author{Oleg P. Sushkov}
\affiliation{School of Physics, University of New South Wales, Kensington, NSW 2052, Australia}

\date{\today}

\title{Exciton condensation from level repulsion: application to bilayer graphene}

\begin{abstract}
Exciton condensation in semiconductors and semimetals has long been predicted but remains elusive. In a semiconductor, condensation occurs when the exciton binding energy matches the band gap. This binding energy results from a balance between Coulomb attraction, which enhances it, and kinetic energy, which suppresses it. However, reducing kinetic energy typically increases screening, weakening Coulomb attraction. Empirically, in most candidate materials, the binding energy remains below the band gap, with few external parameters capable of altering this balance. Here, we propose an in-plane electric field as a control parameter. This field induces hybridisation between even- and odd-parity excitons, and the resulting level repulsion effectively enhances binding energy. We argue that this mechanism is generic to excitons in semiconductors and illustrate it with a model of biased bilayer graphene. Bilayer graphene is chosen since it has a tunable band gap, making it an excitonic condensate candidate and moreover, the Zener tunnelling rate contains, in addition to the usual exponential decay, a non-standard oscillating component -- thanks to details of the electron dispersion. Analogous to quantum oscillations, we propose that Fourier spectrum of the current-voltage data allows for a novel test of exciton condensation. Finally, we show that the is a large excitonic gap to critical temperature ratio -- a clear prediction for STM studies. 
\end{abstract}

\maketitle

\section{Introduction}

Exciton condensates in two-dimensional (2D) materials have long been predicted to exhibit remarkable quantum phenomena, such as dissipationless transport \cite{Lozovik1975, pogrebinskii1977mutual, blatt1962bose, kellogg2004vanishing, su2008make} and topological order \cite{Wang2019, Varsano2020, Perfetto2020, Sun2021, Liu2021topo, scammell2022chiral}. These features make them promising candidates for future quantum technologies. Despite sustained theoretical and experimental effort, however, a robust and unambiguous realisation of zero-field exciton condensation has remained out-of-reach.

Graphene-based systems have emerged as a promising platform in this pursuit. The near particle-hole symmetry of graphene makes it an appealing host for excitonic pairing. Various theoretical proposals have explored condensation in bilayer graphene with either AB stacking under no bias \cite{Nandkishore2010, Song2012, Apinyan2016} with bias \cite{ScammellSushkovPRR2023} or AA stacking with bias \cite{Akzyanov2014, Apinyan2021}, yet there has been no corresponding experimental confirmation. 


In this work, we focus on biased bilayer graphene (Bblg). At small interlayer bias, i.e., a $c$-axis electric field, Bblg develops a tunable gap in its electronic spectrum ~\cite{Zhang2009} and with dual gating the carrier density can be independently controlled~\cite{Taychatanapat2010,engdahl2025}. This electric field tunability has been argued to make Bblg an ideal platform for realisation of exciton condensation \cite{ScammellSushkovPRR2023}. Although the work of Ref. \cite{ScammellSushkovPRR2023} predicts condensation of the $s$-wave exciton, it was recognised that observing this condensate experimentally is challenging due to its extreme sensitivity to screening. Screening arises from both intrinsic and extrinsic sources: dielectric environment, metallic gates, and conduction-band thermal population. In particular, thermal activation of carriers significantly enhances metallic screening and inhibits condensation. Overcoming these effects requires aggressive device engineering—for example, using suspended Bblg to eliminate dielectric screening and placing metallic gates at distances exceeding $20$ nm—conditions that are difficult to realise in standard experimental setups.


This motivates a natural question: can alternative, more accessible control parameters be used to facilitate exciton condensation?

We answer this in the affirmative by showing that an {\it in-plane} electric field offers a powerful and previously underexplored route. The core idea is that such a field induces a dipole matrix element between the $s$- and $p$-wave excitonic states. This coupling results in level repulsion, which lowers the energy of one of the hybridised branches. The resulting enhancement in binding energy strengthens the condensation transition. In this sense, the mechanism resembles the static limit of exciton-polariton physics, but is realised here without optical pumping. The use of an in-plane field thus provides a practical, tunable handle for promoting exciton condensation in conventional device geometries.

Our modelling yields concrete experimental signatures—such as a change in the oscillatory frequency of the tunnelling current and a unique temperature dependence of the band gap shift detectable via STM —that directly probe the presence of the exciton condensate.

\begin{figure*}[t!]
    \centering
\includegraphics[width=0.975\linewidth]{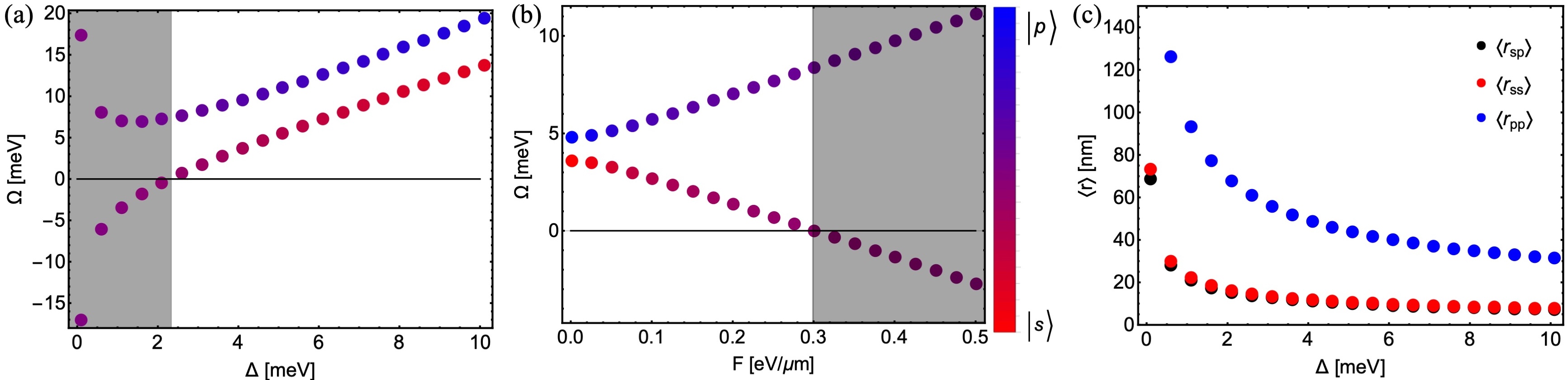}\vspace{0.1cm}
    \caption{$s$- and $p$-wave excitons. (a) Excitation energy as a function of bias $\Delta$ at fixed $F =0.25$ eV/$\mu$m. (b) At fixed bias $\Delta = 2.6$ meV and as a function of in-plane field. In (a) and (b) colors indicate the mixing of the $s$- and $p$-wave eigenstates and gray regions indicate breakdown of LSE and exciton condensation. (c) Characteristic radii vs $\Delta$ at zero electric field $F=0$. Everywhere we have fixed $\varepsilon_r=3.9$ and $d=20$nm.}
    \label{fig:level_rep}
\end{figure*}

The remainder of the paper is structured as follows. Section~\ref{sec:model} presents the modelling of excitonic bound states via the Lippmann–Schwinger equation. The required ingredients are a minimal Hamiltonian for the conduction and valence bands of biased bilayer graphene (Bblg), along with the Coulomb interaction, which is treated using the RPA-screened potential. Section~\ref{sec:zener} introduces an in-plane electric field and examines its effects: (i) modification of the excitonic spectrum, and (ii) the resulting Zener tunnelling rate. Section~\ref{sec:QFT} focuses on the excitonic condensate phase, which emerges as the bound state energy vanishes. An effective quantum field theory is developed to describe the excitonic order parameter, fluctuation corrections, the excitation spectrum, and the critical temperature. Section~\ref{sec:disc} discusses the feasibility of applying in-plane fields, outlines the broader implications of our results, and suggests experimental probes. Appendix~\ref{sec:app1} provides further details on the modelling, including RPA screening and vertex form factors relevant to the bound state problem. Appendix~\ref{sec:app2} presents the derivation of the effective field theory for the excitonic condensate.

\section{Modeling Bblg and Excitonic Bound States}\label{sec:model}
\subsection{Bblg Hamiltonian}

The low-energy single-particle Hamiltonian of Bblg takes the form \cite{McCann2013}
\begin{eqnarray}
\label{H1}
  H=\left(
  \begin{array}{c c}
    \Delta -\mu & -\frac{k_-^2}{2m}\\
    -\frac{k_+^2}{2m} & -\Delta -\mu
  \end{array}
  \right).
\end{eqnarray}
The Hamiltonian is written in the $\{A_1, B_2\}$ orbital basis, where $A,B$ label graphene sublattices and subscripts $1,2$ denote the top and bottom layers, respectively. Here, $k_\pm = \tau k_x \pm i k_y$ with in-plane momentum $\bm{k}$, valley index $\tau = \pm 1$, and effective mass $m \approx 0.032m_e$. The parameter $\Delta$ sets the band gap and is proportional to an applied electric field along the $z$-axis, i.e., $\Delta \propto F_z$. Throughout this work, we set the chemical potential to zero (half-filling), $\mu = 0$.

The resulting single-particle spectrum consists of particle-hole symmetric conduction and valence bands, i.e. $\varepsilon^c_{\bm k}=-\varepsilon^v_{\bm k}=\varepsilon_{\bm k}$, with
\begin{align}
\varepsilon_{\bm k} = \sqrt{\frac{\bm{k}^4}{(2m)^2} + \Delta^2}.
\end{align}
The eigenstates are denoted $|\psi^{(\pm)}_{\bm k,\tau}\rangle$, where the $\pm$ labels refer to conduction and valence band states, respectively.

\subsection{RPA Coulomb}

The interaction responsible for electron–hole attraction is the screened Coulomb potential, given by
\begin{align}
\label{Int}
V_{\bm q, i\xi,T} = -\frac{2\pi \mathrm{e}^2}{\varepsilon_r q/\Upsilon_q - 2\pi \mathrm{e}^2 \Pi({\bm q}, i\xi, T)}.
\end{align}
Here, $\bm q$ denotes the momentum transfer, $\xi$ the imaginary frequency, and $T$ the temperature. The components of this expression are as follows:
(i) the factor $\Upsilon_q = \tanh(q d)$ captures screening due to metallic gates positioned a distance $d$ above and below the bilayer graphene (Bblg); 
(ii) $\varepsilon_r$ represents the effective dielectric constant of the medium between Bblg and the gates; and 
(iii) $\Pi({\bm q}, i\xi, T)$ is the finite-temperature polarisation function of Bblg, discussed in detail in Appendix~\ref{sec:app1}. Finally, we distinguish charge $\mathrm{e}$ from Euler's $e$.

Throughout, we adopt $\varepsilon_r = 3.9$ and interlayer separation $d = 20\,\mathrm{nm}$, corresponding to bilayer graphene encapsulated in hBN with metallic gates placed symmetrically at $20\,\mathrm{nm}$ above and below the structure.

\begin{figure*}[t!]
\includegraphics[width=0.975\linewidth]{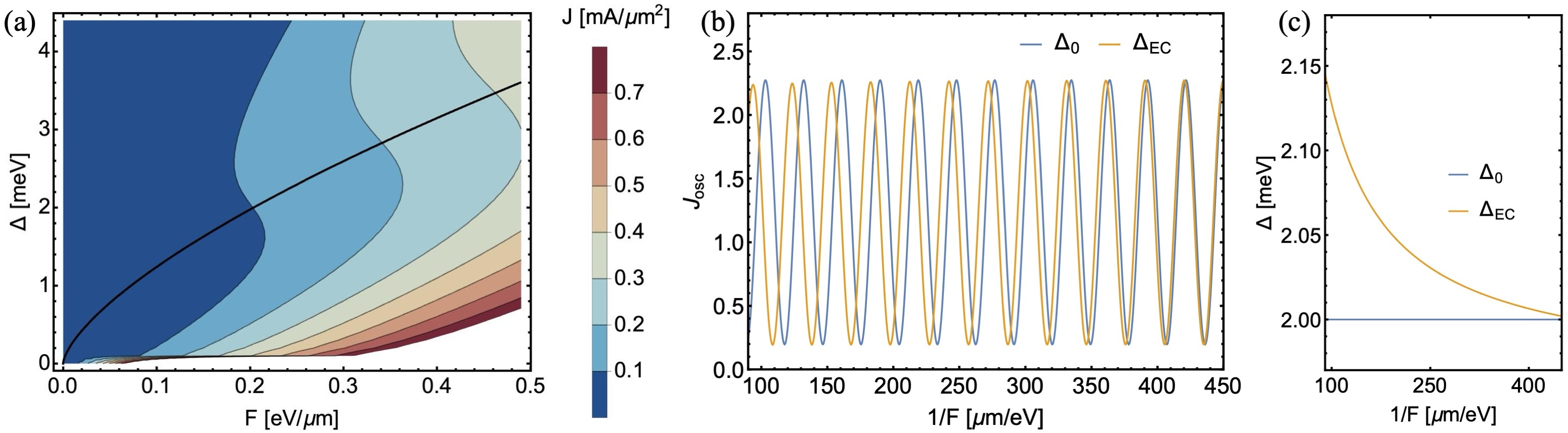}
    \caption{Zener tunnelling characteristics. (a) Contours of the pair-production rate (converted to current per unit area), with phase boundary $\Delta_c(F)$ (black curve) computed from LSE; below the curve is the excitonic condensate. Oscillatory features are evident. (b) Oscillatory component of the current vs $1/F$. The blue curve corresponds to a fixed band gap $\Delta=\Delta_0 = 2$ meV, while the yellow curve corresponds to a field-dependent band gap $\Delta=\Delta_\text{EC}(F)$, due to the exciton condensate, which is plotted in (c). }
    \label{fig:Zener}
\end{figure*}

\subsection{Lippmann-Schwinger Equation}

Assuming an instantaneous interaction—i.e., $V_{{\bm k-\bm k'},T} \equiv V_{\bm q, i\xi=0, T}$ as defined in Eq.~\eqref{Int}—the exciton is described by a wave function obeying the Lippmann-Schwinger equation (LSE),
\begin{align}
  \label{LSE}
(\Omega-2\varepsilon_{\bm k} ) \Psi_{\bm k} =\int \frac{d^2k'}{(2\pi)^2}
         V_{{\bm k-\bm k'}}Z_{\bm {k,k'}}^{\tau',\tau}\tanh\left(\frac{\varepsilon_{\bm k'} }{2T}\right)\Psi_{\bm k'}.
\end{align}
Here, $\Omega$ is the exciton energy at zero total momentum, and the form factor is defined as $Z_{\bm k_1,\bm k_2}^{\tau',\tau} = \langle\psi^{(-)}_{\bm {k}_2,\tau'}|\psi^{(-)}_{\bm {k}_1,\tau'}\rangle\langle\psi^{(+)}_{\bm {k}_1,\tau}|\psi^{(+)}_{\bm k_2,\tau}\rangle$. The LSE is a linear eigenvalue problem that can be readily solved to obtain the eigenvalue $\Omega$ and corresponding eigenfunction $\Psi_{\bm k}$. The equation is well defined for $\Omega \geq 0$. When $\Omega < 0$, the system enters the exciton condensate phase. (In Section~\ref{sec:QFT}, we employ an appropriate method for handling the condensate regime with $\Omega < 0$.) Equivalently, condensation occurs when the exciton binding energy, defined as $\varepsilon_b = 2\Delta - \Omega$, exceeds the band gap, i.e., when $\varepsilon_b > 2\Delta$.

This modelling approach—solving the LSE \eqref{LSE} using the effective Bblg \eqref{H1} Hamiltonian and the RPA-screened Coulomb interaction \eqref{Int}—has previously been shown \cite{ScammellSushkov2022} to accurately reproduce the spectrum of $s$- and $p$-wave excitonic bound states observed experimentally \cite{Ju2017}.

\section{Influence of In-Plane Field}\label{sec:zener}

\subsection{Level Repulsion in the Excitonic Spectrum}

We investigate the influence of an in-plane electric field ${\bm F}$ on the excitonic spectrum of biased bilayer graphene. We orient ${\bm F}$ along the $x$-axis and absorb electric charge $(-\mathrm{e})$, such that the corresponding potential is ${\bm F} \cdot \bm{r} = Fx$. 

We begin by computing the excitonic spectrum in the absence of the in-plane field using the two-body LSE~\eqref{LSE}. We then evaluate the matrix element $\braket{s|Fx|p_x}$, and define $r_{sp} \equiv |\braket{s|x|p_x}|$ to characterise the electric dipole coupling between the $s$- and $p$-like excitons. Here $\ket{s}, \ket{p_x}$ refer to LSE eigenstates, $\Psi_{\bm k}$ of Eq.~\eqref{LSE}, yet decomposed into the angular momentum channels. The angular decomposition is detailed in the Appendix \ref{App:rsp}. We denote the corresponding eigenvalues $\Omega_s$, $\Omega_p$. 

To capture the level repulsion induced by the electric field, we employ an effective two-level Hamiltonian:
\begin{align}
H_\text{eff} = \begin{pmatrix}
\Omega_s & F r_{sp} \\
F r_{sp} & \Omega_p
\end{pmatrix}.
\end{align}
The eigenvalues of $H_\text{eff}$ exhibit the desired level-repuslion,
\begin{align}
    \Omega_\pm & =\frac{1}{2} \left(\Omega _p+\Omega_s\right)\pm \sqrt{(F r_{sp})^2+\frac{1}{4}(\Omega_s-\Omega_p)^2},
\end{align}
with $\Omega_-$ the lower-in-energy branch. 
From this bound-state perspective, one sees that for $F > \sqrt{\Omega_s \Omega_p}/r_{sp}$, the lower branch becomes unstable $\Omega_-<0$, indicating condensation of the hybridised $s$-$p$ mode. 

Figures~\ref{fig:level_rep}(a) and (b) display the evolution of the excitonic spectrum with increasing $F$, illustrating the expected level repulsion. To further clarify the role of electric dipole coupling, Fig.~\ref{fig:level_rep}(c) presents the computed $r_{sp}$ alongside the root-mean-square radii $\sqrt{\braket{s|r^2|s}}$ and $\sqrt{\braket{p|r^2|p}}$ vs band gap $\Delta$ at zero field $F=0$. 

One can also compute the $(\Delta, F)$ phase boundary within this LSE scheme: it corresponds to the vanishing of the lowest eigenvalue, i.e. $\Omega_-(\Delta_c,F)=0$. The phase boundary $\Delta_c(F)$ is superimposed on Fig.~\ref{fig:Zener}(a).

\subsection{Zener Tunnelling}

An in-plane electric field also induces Zener tunnelling between the valence and conduction bands. Building on results from a companion work \cite{ScammellSushkov2025_Zener} (see also earlier work \cite{LevitovPNAS2011}), we compute the corresponding tunnelling current per unit area,
\begin{align}
  \label{dn}
  \notag &J=
   \frac{\mathrm{e}(m\Delta^2)}{2\pi(\pi\beta_1)^{1/2}}\left(\frac{F^2}{m\Delta^3}\right)^{3/4} \left[1-\frac{\cos(S-\frac{\pi}{8})}{2^{1/4}}\right]e^{-S},\\
   &S=2\beta_0\left(\frac{m\Delta^3}{F^2}\right)^{1/2}, \quad \beta_0=1.748.
\end{align}
Figure~\ref{fig:Zener}(a) shows the computed tunnelling current across the phase diagram—that is, as a function of both $F$ and $\Delta$.  As mentioned above, the excitonic phase boundary $\Delta_c(F)$ separating the normal state from the exciton condensate is superimposed. Accordingly, to minimise tunnelling current and sit within the excitonic condensate, an optimal range is $F\lesssim 0.2$ eV/$\mu$m and with band gap $\Delta\lesssim 2$ meV. This analysis neglects the influence of the excitonic enhanced band gap on the tunnelling current, which would further exponentially suppress the current. Hence, Fig~\ref{fig:Zener}(a) should be understood as the upper bound on the tunnelling current.

An interesting feature of the current is the non-monotonic dependence on $F$. To examine the non-monotonic behaviour further, we consider the oscillatory component of the current, defined as $J_\text{osc} \equiv J\, e^{S}/F^{3/2}$. For a field-independent gap $\Delta$, the function $J_\text{osc}$ oscillates periodically in $1/F$. Figure~\ref{fig:Zener}(b) plots $J_\text{osc}$ as a function of $F$ for both a constant band gap $\Delta = \Delta_0$ and a field-dependent, exciton-corrected gap $\Delta = \Delta_\text{EC}(F)$. A weak $F$-dependence in $\Delta$ manifests as a clear shift in the oscillation frequency, thereby breaking the $1/F$ periodicity.

Crucially, within the condensate phase, the band gap $\Delta$ itself depends on $F$—a central claim of this work, elaborated in Section~\ref{sec:QFT}. We propose that the resulting shift in oscillation frequency provides a robust signature of exciton condensation. We note that this approach is conceptually similar to quantum oscillation techniques, where magnetoresistance oscillations as a function of inverse magnetic field reveal the area of the Fermi surface \cite{Shoenberg1984}.

\section{Properties of the Exciton Condensate: Field Theory}\label{sec:QFT}
We now analyse the field-induced exciton condensate phase. Using a low-energy effective field theory built from the $s$- and $p$-wave excitonic states, we characterise the excitation spectrum, symmetry-breaking effects, thermal stability, and experimental observables.

\begin{figure*}[t!]
    \centering
\includegraphics[width=0.8\linewidth]{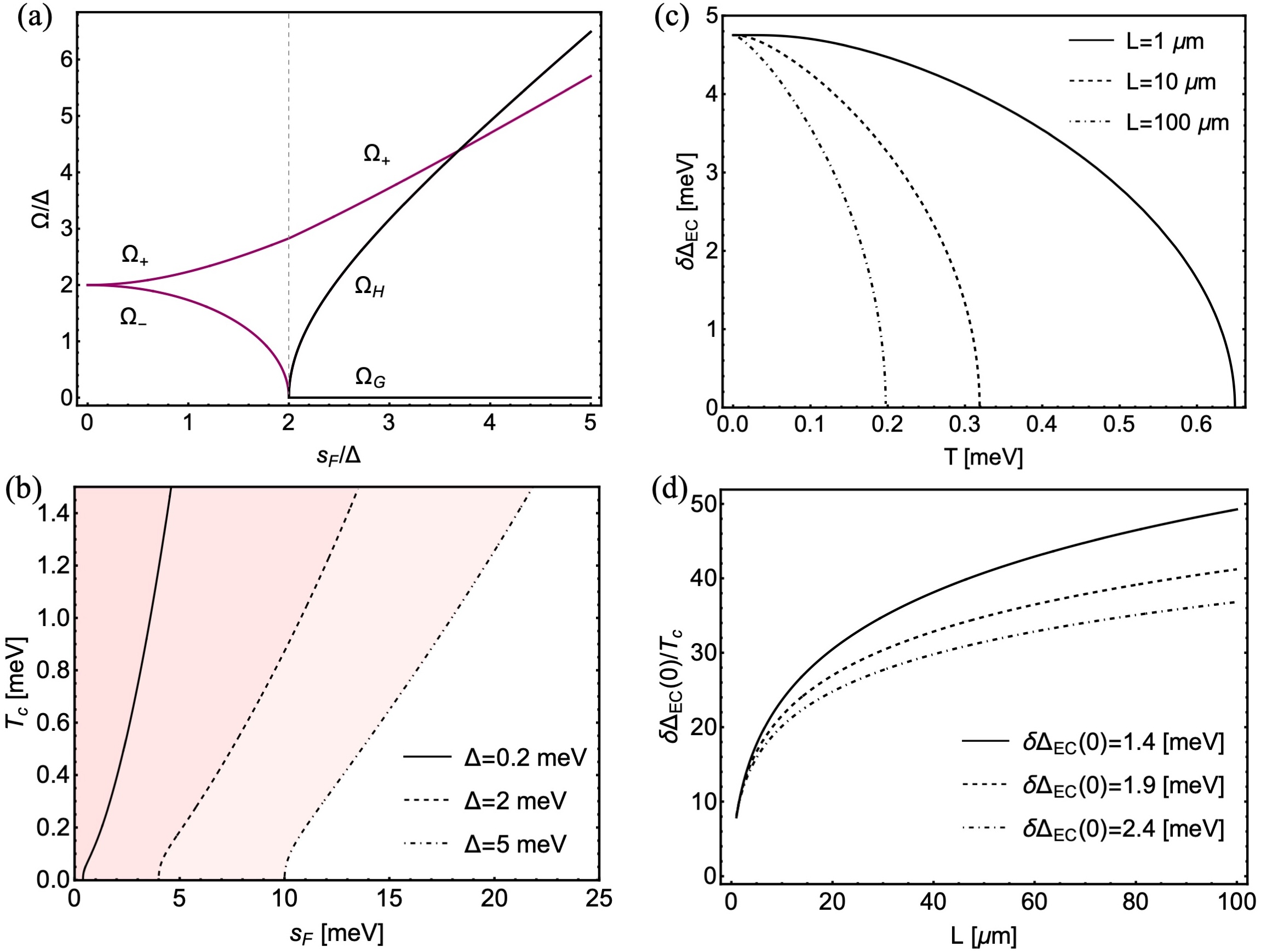}
    \caption{Exciton Condensate: (a) Evolution of modes across the phase transition. (b) $(s_F, T)$ phase diagram, parametrised by different $\Delta$, at fixed $L=10\mu$m. (c) Melting of the excitonic gap due to Goldstone fluctuations, with Goldstone gap set by system size $\Omega_G=c\pi/L$. (d) Ratio of the zero-temperature excitonic gap correction to the critical temperature, $\delta\Delta_\text{EC}(0)/T_c$, vs $L$, parametrised by the zero-temperature gap correction $\delta\Delta_\text{EC}(0)$ with a fixed $\Delta=2$meV.}
    \label{fig:EC}
\end{figure*}

\subsection{Effective Quantum Field Theory}

We model the excitonic sector using a Euclidean field theory involving $s$- and $p$-wave bound states:
\begin{align}
\label{Lfull}
\notag &{\cal L}= \\
\notag &\begin{pmatrix} \Phi_s^* \\ \Phi_{p_x}^* \end{pmatrix}^T
\begin{pmatrix}
-\partial_\tau^2 - c^2 \nabla^2 + s^2 & s_F^2 \\
s_F^2 & -\partial_\tau^2 - c^2 \nabla^2 + s^2
\end{pmatrix}
\begin{pmatrix} \Phi_s \\ \Phi_{p_x} \end{pmatrix}\\
&+ \lambda \left( |\Phi_s|^4 + 6 |\Phi_s|^2 |\Phi_{p_x}|^2 + \tfrac{3}{2} |\Phi_{p_x}|^4 \right).
\end{align}
Here $s=\Omega=2\Delta-\varepsilon_b$ is the exciton oscillator strength, and $s_F\propto F$ quantifies the electric-field-induced hybridisation. Condensation occurs when $s_F^2 > s^2$. Hereon we treat $s_F$ as a tuning parameter. The exciton velocity and interaction strength are
\begin{align}
c= 1.15\sqrt{\frac{\Delta}{2m}}, \quad \lambda = 50.3 \frac{\Delta^2}{2m}.
\end{align}
 The procedure for computing these numerical coefficients is detailed in  Appendix~\ref{sec:app2}.

 We now turn to the excitation spectrum of this theory and the consequences of symmetry breaking.

\subsection{Excitation Spectrum and Symmetry Breaking}

In the condensed phase, the lowest mode is the hybrid combination $\Phi_- = (\Phi_s - \Phi_{p_x})/\sqrt{2}$. The effective theory becomes
\begin{align}
\label{Leff}
{\cal L}_-&=\Phi_-^*\left(-\partial_\tau^2 - c^2 \nabla^2 + (s^2-s_F^2)\right)\Phi_- + \frac{17}{8}\lambda |\Phi_-|^4.
\end{align}
This model exhibits spontaneous breaking of U(1) symmetry when $s_F^2>s^2$, giving rise to two collective excitations
\begin{align}
\omega_{\bm k,H}^2 = c^2\bm k^2 + \Omega_H^2, \quad \omega_{\bm k,G}^2 = c^2\bm k^2,
\end{align}
where $\Omega_H^2 = 2|s^2 - s_F^2|$ is the amplitude (Higgs) gap, and the Goldstone mode is gapless in the thermodynamic limit. The mass gap of the each mode of \eqref{Lfull} vs electric field tuning parameter $s_F$ is shown in Fig.~\ref{fig:EC}(a). 

Physically, we expect the Higgs mode to lie at the two-particle threshold, as per Refs \cite{Volkov1973,Gurarie2009,Tokimoto2019}, rendering it heavily damped and challenging to detect. 
 
Having characterised the ground state and its excitations, we now examine the role of thermal fluctuations and finite-size effects.

\subsection{Thermal Fluctuations and Finite-Size Critical Temperature}

Due to the Mermin-Wagner theorem, true long-range order is forbidden at any $T>0$ in infinite 2D systems with continuous symmetry. However, in finite-size systems, a pseudo-gap opens for the Goldstone mode: min$[\omega_{\bm k,G}]\sim c\pi/L$. We therefore compute the condensate stability as a function of temperature for a typical system size $L\sim 10~\mu$m.

The order parameter $\nu = \braket{\Phi_-}$ satisfies at $T=0$
\begin{align}
\nu^2 = \frac{4|s^2 - s_F^2|}{17 \lambda}.
\end{align}
At finite $T$, thermal occupation of collective modes reduces $\nu$. A perturbative expansion in $\lambda$ yields \cite{Scammell2018},
\begin{align}
\label{nu_T}
\nu^2(T) = \nu^2 - \sum_{\bm k} \frac{n(\omega_{\bm k,G})}{\omega_{\bm k,G}} - 3\sum_{\bm k} \frac{n(\omega_{\bm k,H})}{\omega_{\bm k,H}}.
\end{align}
The Goldstone contribution dominates this reduction and defines the critical temperature $T_c(L)$ via $\nu^2(T_c)=0$. Plots of $T_c$ vs $s_F$ are shown in Fig.~\ref{fig:EC}(b) -- there we keep $L$ fixed and parametrise by $\Delta$. 

It is important to address the role of phase fluctuations. Since the fermionic gap depends on the magnitude of the excitonic order parameter $|\nu|$, and not its phase, one might ask how phase fluctuations impact the picture. In particular, if the amplitude were fixed, $|\nu| \approx$ constant, yet phase fluctuations rendered the spatial/temporal average zero, then the influence on the fermionic spectrum would be unclear and require more care. However, in the present situation, both phase and amplitude fluctuations are present; the amplitude modes become gapless as $T \to T_c$, and with this, the magnitude of the order parameter vanishes. To see this, we reiterate that a nonzero order parameter $\nu \neq 0$ arises when the mass term $\delta s^2 = s_F^2 - s^2$ in Eq.~\eqref{Leff} is positive. Thermal renormalisation is such that $\delta s^2(T) \to 0$ as $T \to T_c$, (as per e.g. \cite{Scammell2018}) and accordingly, the amplitude vanishes $|\nu(T_c)| = 0$.

\subsection{Condensate-Induced Gap Shift and Experimental Signatures}

The key observable is the shift in the electronic band gap due to condensation. In momentum space, the excitonic mean-field $\phi_-(\bm k)$ modifies the bare gap
\begin{align}
\Delta^2_{\bm k} = \Delta^2 + |\phi_-(\bm k)|^2.
\end{align}
Here $\phi_-(\bm k) = (\Xi_{s}(\bm k)\Phi_s-\Xi_{p}(\bm k)\Phi_{p_x})/\sqrt{2 \chi_\perp}$, where $\chi_\perp=0.064\times (2m/\Delta^2)$ is a dimensionful scaling parameter (derived in the Appendix \ref{sec:app2}), which converts the field $\phi_-(\bm k)$ to units of energy, and $\Xi_{s/p}(\bm k)$ are $\bm k$-space basis functions transforming as $s$- and $p$-wave irreducible representations; up to normalisation they are $\Xi_{s}(\bm k)\sim1, \ \Xi_{p}(\bm k)\sim k_x$.  At $\bm k = 0$, this gives the gap shift 
\begin{align}
\delta\Delta_\text{EC} \equiv |\phi_-(\bm k=\bm 0)| = \nu(T)/\sqrt{2\chi_\perp},
\end{align}
which can be directly probed in STM or transport. In Fig.\ref{fig:EC}(c) we show the temperature dependence of $\delta\Delta_\text{EC}$ for several values of $L$.

Finally, in Fig.~\ref{fig:EC}(d), we plot the ratio of the zero-temperature excitonic gap correction to the critical temperature, $\delta\Delta_\text{EC}(0)/T_c$, as a function of system size $L$.
 We conclude that the large ratio $\delta\Delta_\text{EC}(T=0)/T_c$ is a hallmark of the phase mode, and hence of true exciton condensation. This is testable via STM.

\section{Discussion}\label{sec:disc}

{\it Summary.} We revisit the longstanding problem of exciton condensation from a semiconducting ground state. Although Coulomb attraction between conduction and valence bands is the canonical driving mechanism, the lack of robust experimental realisations suggests that this interaction alone is insufficient. In this work, we identify a new control parameter: level repulsion induced by an in-plane electric field. We demonstrate that this additional handle can significantly enhance the effective exciton binding energy and, in turn, drive a condensation transition.

Our analysis shows that, above a critical in-plane field strength, the excitonic spectrum becomes unstable—signalling the onset of condensation. However, the same electric field also enables interband Zener tunnelling, which can potentially undermine the condensate. To address this, we compute the Zener tunnelling current, as derived in Ref.~\cite{ScammellSushkov2025_Zener}, to identify an optimal regime in biased bilayer graphene (Bblg) that simultaneously supports condensation while suppressing current.

Moreover, the tunnelling current exhibits non-monotonic dependence on electric field due to an oscillatory prefactor. The oscillation period is sensitive to excitonic corrections to the band gap, offering a new means to probe condensation in Bblg: by measuring the tunnelling current as a function of field and extracting its Fourier components, one obtains an experimental signature analogous to quantum oscillations used to extract Fermi surface properties. This is made possible by the oscillatory nature of Zener tunnelling in Bblg which, as identified in Ref.~\cite{LevitovPNAS2011}, is a direct consequence of its unusual band structure. By contrast, gapped Dirac or Schr\"odinger-type semiconductors exhibit simple, monotonic Zener behaviour.

To describe the excitonic condensate phase, we construct an effective quantum field theory, which reveals several distinctive features. In particular, we find that the ratio of the excitonic band gap correction to the critical temperature is anomalously large (much greater than unity) and system-size dependent. In the thermodynamic limit, this ratio diverges, reflecting the Mermin-Wagner theorem. We stress that this provides a sharp, testable prediction for the excitonic condensate; one that can be accessed via, e.g., temperature-dependent STM measurements of the band edge.

{\it Feasibility.} The proposed mechanism relies on in-plane electric fields to enhance condensation via hybridisation. Experimental work on transition metal dichalcogenides has demonstrated field strengths up to $10$ eV/$\mu$m \cite{Zhu_TMD_Field}, which exceed the critical field required in Bblg by roughly an order of magnitude. This suggests that the necessary conditions are experimentally accessible using standard techniques.

{\it Outlook.} Although we have focused on Bblg, the underlying mechanism is broadly applicable. Any semiconductor with low-lying excitonic states may benefit from field-induced level repulsion to stabilise condensation. In contrast, materials in the BCS regime—where the normal state is (semi)metallic—may be less suitable, as in-plane fields drive dissipative conduction. These considerations suggest new routes for engineering excitonic condensates across a wide range of material platforms.\\

\section{Acknowledgements}
We thank Bent Weber and Yande Que for insightful discussions on related ideas.


%

\appendix

\section{Excitonic Bound States}\label{sec:app1}
\subsection{Hamiltonian and wavefunctions}
The Hamiltonian and wavefunctions are
\begin{align}
\notag H&=-\frac{(\tau k_x \mp i k_y)^2}{2m}\sigma_\pm + \Delta\sigma_z\\
\notag |\psi^{(+)}_{\bm k,\tau}\rangle&=\frac{1}{\sqrt{\frac{(\varepsilon_{\bm k}-\Delta)^2}{k^4/(4m^2)}+1}}\begin{pmatrix} -1 \\ \frac{\varepsilon_{\bm k}-\Delta}{k^2/(2m)}e^{2i\tau \theta_{\bm k}} \end{pmatrix},\\
\notag |\psi^{(-)}_{\bm k,\tau}\rangle&= \frac{1}{\sqrt{\frac{(\varepsilon_{\bm k}-\Delta)^2}{k^4/(4m^2)}+1}}\begin{pmatrix} \frac{\varepsilon_{\bm k}-\Delta}{k^2/(2m)}e^{-2i\tau \theta_{\bm k}} \\ 1 \end{pmatrix}.
\end{align}

\subsection{Polarization Operator}\label{polop_supp}
The polarization operator is given by,
\begin{align}
         \Pi({\bm q},i\xi,T)&=4 \sum_{\mu,\nu=\pm}  \int \frac{d^2 k}{(2\pi)^2} \frac{\left(f_{\varepsilon^\mu_{\bm k}}-f_{\varepsilon^\nu_{\bm {k+q}}}\right)}
    {i\xi + \varepsilon^\mu_{\bm k}-\varepsilon^\nu_{\bm {k+q}}}F^{\mu\nu}_{\bm {k,k+q}},
\end{align}
with form factors $F^{\mu\nu}_{\bm {k,k+q}}=|\langle\psi^{(\mu)}_{\bm {k+q},\tau}|\psi^{(\nu)}_{\bm k,\tau}
\rangle|^2$, and indices $\mu,\nu=\pm$.
Using the relation $-\varepsilon^-_{\bm k}=\varepsilon^+_{\bm k}\equiv \varepsilon_{\bm k}$ and $F^{--}_{\bm {k,k+q}}=F^{++}_{\bm {k,k+q}}$, $F^{-+}_{\bm {k,k+q}}=F^{+-}_{\bm {k,k+q}}$, we rearrange into the form 
\begin{gather}
  \label{polT}
       \notag  \Pi({\bm q},i\xi,T)=8 \int \frac{d^2 k}{(2\pi)^2}\bigg[ \frac{(\varepsilon_{\bm k}-\varepsilon_{\bm {k+q}})\left(f_{\varepsilon_{\bm k}}-f_{\varepsilon_{\bm {k+q}}}\right)}
    {\xi^2 + (\varepsilon_{\bm k}-\varepsilon_{\bm {k+q}})^2}F^{++}_{\bm {k,k+q}}\\ + \frac{(\varepsilon_{\bm k}+\varepsilon_{\bm {k+q}})\left(f_{\varepsilon_{\bm k}}+f_{\varepsilon_{\bm {k+q}}}-1\right)}
    {\xi^2 + (\varepsilon_{\bm k}+\varepsilon_{\bm {k+q}})^2}F^{+-}_{\bm {k,k+q}} \bigg].
\end{gather}

\subsection{Vertex Form Factors}

We consider a Coulomb interaction of the form
\begin{align}
&H_{int} =\\
\notag &\sum_{\bm k_1,\bm k_2,\bm k_3}\sum_{\tau,\tau'}V_{\bm k_2-\bm k_1}\psi^\dag_{\bm k_1,\tau} \psi^\dag_{\bm k_2,\tau'} \psi_{\bm k_3,\tau'} \psi_{\bm k_1+\bm k_2 -\bm k_3,\tau},
\end{align}
where the $\psi^\dag_{\bm k,\tau}$ are electron creation operators. Neglecting any processes involving valley exchange. To proceed, we transform to the band basis via
\begin{align}
\notag \psi_{\bm k,\tau} &= {\cal U}_{\bm k,\tau}\begin{pmatrix} c_{\bm k,\tau}\\ v_{\bm k,\tau} \end{pmatrix},\quad {\cal U}_{\bm k,\tau}=(|\psi^{(+)}_{\bm k,\tau}\rangle, |\psi^{(-)}_{\bm k,\tau}\rangle),
\end{align}
where $c_{\bm k,\tau}$ and $v_{\bm k,\tau}$ are the annihilation operators for conduction and valence states, respectively. Focusing on the excitonic channel, denoted $H_{int}^X$, the interaction simplifies to
\begin{align}
\notag &H_{int}^X =\sum_{\bm k_1,\bm k_2}\sum_{\tau,\tau'}V(\bm k_2-\bm k_1)c^\dag_{\bm k_1,\tau} c_{\bm k_2,\tau} v^\dag_{\bm k_2,\tau'} v_{\bm k_1,\tau'}\\
&\times \langle\psi^{(-)}_{\bm {p}_2,\tau'}|\psi^{(-)}_{\bm {p}_1,\tau'}\rangle\langle\psi^{(+)}_{\bm {p}_1,\tau}|\psi^{(+)}_{\bm k_2,\tau}\rangle.
\end{align}
We define the form factor
$
Z_{\bm k_1,\bm k_2}^{\tau',\tau} = \langle\psi^{(-)}_{\bm {p}_2,\tau'}|\psi^{(-)}_{\bm {p}_1,\tau'}\rangle\langle\psi^{(+)}_{\bm {p}_1,\tau}|\psi^{(+)}_{\bm k_2,\tau}\rangle.$

\subsection{Details of LSE eigenvalue problem}\label{App:rsp}
Considering the Lippmann-Schwinger
equation presented in the main text
\begin{align}
  \label{app_LSE}
(\Omega-2\varepsilon_{\bm k} ) \Psi_{\bm k} =\int \frac{d^2k'}{(2\pi)^2}
         V_{{\bm k-\bm k'}}Z_{\bm {k,k'}}^{\tau',\tau}\tanh\left(\frac{\varepsilon_{\bm k'} }{2T}\right)\Psi_{\bm k'},
\end{align}
we use an angular decomposition of the eigenfunctions
\begin{align}
\Psi_{\bm k}& = \sum_{\ell} \frac{2\pi}{\sqrt{k}}\Phi^\ell_{k} e^{i\ell\theta_{\bm k}}.
\end{align}
Normalization is such that $\sum_k |\Phi^\ell_{k}|^2=1$,  which is quite natural when numerically diagonalizing the LSE. In this convention, the matrix element $r_{sp}$ is computed as
\begin{align}
r_{sp} &= |\braket{p_x|i\partial_{k_x}|s}|\\
\notag &= \bigg|\int \frac{d^2k}{(2\pi)^2}\frac{2\pi}{\sqrt{k}}\Phi^{\ell=1}_{k} \cos\theta_{\bm k} \partial_{k_x} \left(\frac{2\pi}{\sqrt{k}}\Phi^{\ell=0}_{k}\right)\bigg|\\
\notag &= \bigg|\int kdkd\theta\frac{1}{\sqrt{k}}\Phi^{\ell=1}_{k} \left(\cos\theta_{\bm k} \frac{\partial_{k}}{\partial_{k_x}}\right) \partial_{k}\left(\frac{1}{\sqrt{k}}\Phi^{\ell=0}_{k}\right)\bigg|\\
\notag &= \bigg|\int d\theta \cos^2\theta \int kdkd\frac{1}{\sqrt{k}}\Phi^{\ell=1}_{k} \partial_{k}\left(\frac{2\pi}{\sqrt{k}}\Phi^{\ell=0}_{k}\right)\bigg|\\
\notag &= \frac{1}{2}\bigg|\int kdkd\frac{1}{\sqrt{k}}\Phi^{\ell=1}_{k} \partial_{k}\left(\frac{1}{\sqrt{k}}\Phi^{\ell=0}_{k}\right)\bigg|.
\end{align}
We compute the derivative numerically via finite difference.

\section{Excitonic Quantum Field Theory}\label{sec:app2}
Considering meanfield excitonic order $\phi_{\bm k}$, we construct the mean field Hamiltonian,
\begin{align}
    \hat{H}_0&=\begin{pmatrix} \varepsilon^c(\bm k) && 0 \\ 0 && \varepsilon^v(\bm k)
    \end{pmatrix}, \quad \hat{H}_\phi=\begin{pmatrix} 0 && \phi_{\bm k} \\ \phi_{\bm k}^* && 0
    \end{pmatrix},\\
    {\cal H}_\text{MF}&= \sum_{\bm k}\psi^\dag \left(\hat{H}_0+\hat{H}_\phi\right) \psi - \frac{1}{V}\phi_{\bm k}^\dag\phi_{\bm k}
\end{align}
with conduction and valence band dispersions $\varepsilon^c(\bm k)=\sqrt{k^4/(2m)^2+\Delta^2}$, $\varepsilon^v(\bm k)=-\varepsilon^c(\bm k)$. 

The free energy is given by
\begin{align}
    &{\cal F}=\Tr\ln\left[i\omega_n - \hat{H}_\text{BdG}\right] - \sum_{\bm k} \frac{1}{V_\text{eff}}\phi_{\bm k}^\dag\phi_{\bm k}.
\end{align}
Expanding the logarithm and keeping terms up to quartic order in $\phi$, we obtain
\begin{align}
\label{Fapp}
    &{\cal F} = \Tr\ln\left[i\omega_n - \hat{H}_0\right] 
    - \sum_{\bm k} \frac{1}{V_\text{eff}} \phi_{\bm k}^\dagger \phi_{\bm k}
    + \frac{1}{2} \Tr \left[ (G \hat{H}_\phi)^2 \right] \notag \\
    &- \frac{1}{3} \Tr \left[ (G \hat{H}_\phi)^3 \right] 
    + \frac{1}{4} \Tr \left[ (G \hat{H}_\phi)^4 \right] 
    + \dots.
\end{align}
Here, \(G(\bm k,i\omega_n) = (i\omega_n - \hat{H}_0(\bm k))^{-1}\) is the unperturbed Green’s function. For later use, we will introduce Pauli matrices $\eta_\mu$ in band basis, such that
\begin{gather}
\label{Geta}
    G(\bm k,i\omega_n) = (i\omega_n - \varepsilon^c(\bm k))(\eta_0+\eta_z)/2\\
    + (i\omega_n - \varepsilon^v(\bm k))(\eta_0-\eta_z)/2.
\end{gather}

Note that we hereon subtract the order parameter independent contribution ${\cal F}_0=\Tr\ln\left[i\omega_n - \hat{H}_0\right]$, but do not employ a new label, i.e. ${\cal F}-{\cal F}_0 \to {\cal F}$. 

\subsection{Angular decomposition}
We perform an angular harmonics decomposition of $\phi_{\bm k}$, keeping $\ell=0,\pm1$ 
\begin{align}
\phi_{\bm k}&= \Xi_\ell(\bm k) \varphi_\ell
\end{align}
with $\Xi_\ell(\bm k)$ transforming with angular momentum $\ell$, e.g. $\Xi_0(\bm k)\sim1$, $\Xi_1(\bm k)\sim k_x+i k_y$, etc, and with appropriate normalisation/dimensionful factors such that they form an orthonormal basis, i.e. $\sum_{\bm k} \Xi_\ell(\bm k)\Xi_{\ell'}(\bm k)=\delta_{\ell\ell'}$. Subsequently, the $\varphi_\ell$ are the momentum-independent order parameters.

The free energy is written as a polynomial in $\varphi_\ell$, and for simplicity we take $V_\text{eff}$ to be the same in both $|\ell|=0,1$ channels, 
\begin{align}
\label{Fell}
    {\cal F}&=\sum_\ell\varphi_\ell^*(\frac{1}{V_\text{eff}} - \Pi)\varphi_\ell  + \gamma U_4[\varphi_\ell]
\end{align}
here $\gamma$ is the quartic interaction coupling constant, and the quartic potential $U_4[\varphi_\ell]$ given by
\begin{align}
 \notag    U_4[\varphi_\ell]&=|\varphi_s|^4 + 
6 |\varphi_s|^2 \left(|\varphi_{p_x}|^2+|\Phi _{p_y}|^2\right)\\
\notag &+\frac{3}{2} \left(
\varphi_{p_x}|^2+|\varphi_{p_y}|^2\right){}^2.
\end{align}
For later convenience, we have gone from chiral to cartesian coordinates, i.e. $\varphi_{0}\to\varphi_{s}$ and $\varphi_{\pm1}\to\varphi_{p_x}\pm i\varphi_{p_y}$. Finally, we specialise to electric field along $x$ and hence only consider mixing of $\varphi_s$ and $\varphi_{p_x}$ excitonic fields. Neglecting the gapped mode $\varphi_{p_y}$, we are left with
\begin{align}
U_4[\varphi_\ell]&=|\varphi_s|^4 + 
6 |\varphi_s|^2 |\varphi_{p_x}|^2
+\frac{3}{2} |
\varphi_{p_x}|^4.
\end{align}

\subsection{Effective QFT}
We construct the effective excitonic QFT by focusing on the lowest-lying $s$- and $p$-wave excitons.  At zero temperature, the two-point susceptibility is given by
\begin{align}
\notag \chi^{(2)}(\bm{q},i\omega) &= \int \frac{d^2k}{(2\pi)^2} \int \frac{d\nu}{2\pi} \\
\notag  &\hspace{-2cm}\text{Tr} \left[
G(\tfrac{\bm{q}}{2} + \bm{k},i\nu + i\omega/2,) \, \eta_+ \,
G(\tfrac{\bm{q}}{2} - \bm{k},i\nu - i\omega/2) \, \eta_-
\right]\\
&=\int \frac{d^2k}{(2\pi)^2} \frac{(\varepsilon^c_{\bm k}-\varepsilon^v_{\bm k})}{(\varepsilon^c_{\bm k+\bm q/2}-\varepsilon^v_{\bm k-\bm q/2})^2 + \omega^2}
\end{align}
where $G(\bm{k},i\nu)$ is defined in  Eq.~\eqref{Geta}, which makes use of the Pauli matrices in band basis $\eta_\mu$. The appearance of $\eta_\pm$ corresponds to the excitonic vertices, i.e. due to pairing of conduction and valence states. The trace is taken over band indices.

The four-point susceptibility governing interactions is, sepcialising to the static, uniform limit, 
\begin{align}
\label{chi4}
\notag \chi^{(4)}(\bm{0},0,) &= \int \frac{d^2k}{(2\pi)^2} \int \frac{d\omega}{2\pi} \,
\text{Tr} \left[
G \eta_+ G \eta_- G \eta_+ G \eta_-
\right]\\
&=\int \frac{d^2k}{(2\pi)^2} \frac{2}{(\varepsilon^c_{\bm k}-\varepsilon^v_{\bm k})^3}.
\end{align}
The four-point susceptibility is needed to define the effective quartic interaction strength, i.e. $\gamma$ appearing in \eqref{Fell}, as we detail below. 

Field theory parameters are determined by susceptibilities computed from fermion loops,
\begin{align}
\notag \chi_\perp &= -\frac{1}{2} \partial^2_\omega \chi^{(2)}(\bm{0}, \omega)\big|_{\omega=0}, \\
\notag \rho_s &= -\frac{1}{2} \partial^2_{\bm{q}} \chi^{(2)}(\bm{q}, 0)\big|_{\bm{q} = \bm{0}}, \\
\notag M^2 &= V_\text{eff}^{-1} - \chi^{(2)}(\bm{0}, 0), \\
\gamma &= \frac{1}{4}\chi^{(4)}(\bm{0}, 0).
\end{align}
Note: we have already accounted for symmetry factors of the diagrams within the free energy expansion. 

Armed with these expressions, the  resulting Euclidean action for excitonic fields $\varphi_\ell$ is
\begin{align}
\notag &{\cal S} 
= \int d\tau\, d^2x \Big[\varphi_s^*(-\chi_\perp \partial_\tau^2 - \rho_s \nabla^2 + M^2) \varphi_s\\
\notag &+\varphi_{p_x}^*(-\chi_\perp \partial_\tau^2 - \rho_s \nabla^2 + M^2) \varphi_{p_x} \\
&+ \lambda \left( |\varphi_s|^4 + 6 |\varphi_s|^2 |\varphi_{p_x}|^2 + \tfrac{3}{2} |\varphi_{p_x}|^4 \right) \Big].
\end{align}
By direct computation, the parameters are
\begin{align}
\chi_\perp=0.01 \frac{2m}{\Delta^2}, \quad \rho_s=0.013 \frac{1}{\Delta}, \quad \gamma=0.005 \frac{2m}{\Delta^2}.
\end{align}
We discuss $M$ next. 

\subsubsection{Rescaled QFT}
It will prove convenient to perform  a rescaling of the field 
\begin{align}
\varphi\to \Phi/\sqrt{\chi_\perp}
\end{align}
under which the physical parameters become
\begin{align}
    c^2 = \rho_s/\chi_\perp, \qquad s^2 = M^2/\chi_\perp, \qquad \lambda = \gamma/\chi^2_\perp.
\end{align}
The convenience enters since, under this rescaling, the oscillator strength $s$ directly corresponds to the bound state energy computed in LSE, i.e. $s=\Omega$. Hence, in this scheme, we have accurately taken into account $V_\text{eff}$, i.e. it follows from the full RPA screened interaction as treated within LSE.

These parameters are computed to be
\begin{align}
c^2=1.3 \frac{\Delta}{2m}, \quad s = \Omega,  \quad \lambda = 50.3 \frac{\Delta^2}{2m}.
\end{align}
\\

\subsubsection{Including electric field}
We now account for the parity-breaking effect of an applied in-plane electric field $F x$. In the effective field theory, this parity-breaking couples the $\Phi_s$ and $\Phi_{p_x}$ fields. We account for this via an phenomenological parameter $s_F$, which in our currency is an field-mixing energy scale associated with the electric field. The conversion between $F$ and $s_F$ involves the parameter $r_{sp}$; in the QFT analysis, we don't repeat the computation of $r_{sp}$, but instead treat $s_F$ itself as the control parameter. 

The resulting Euclidean action is
\begin{align}
\notag &{\cal S} 
= \int d\tau\, d^2x \Big[\\
\notag &\begin{pmatrix} \Phi_s^* & \Phi_{p_x}^* \end{pmatrix}
\begin{pmatrix}
-\partial_\tau^2 - c^2 \nabla^2 + s^2 & s_F^2 \\
s_F^2 & -\partial_\tau^2 - c^2 \nabla^2 + s^2
\end{pmatrix}
\begin{pmatrix} \Phi_s \\ \Phi_{p_x} \end{pmatrix}\\
&+ \lambda \left( |\Phi_s|^4 + 6 |\Phi_s|^2 |\Phi_{p_x}|^2 + \tfrac{3}{2} |\Phi_{p_x}|^4 \right) \Big].
\end{align}
We use this action in the main text.

\end{document}